\documentclass{emulateapj}
\usepackage{color}
\slugcomment{To be published by the Astrophysical Journal}

\shorttitle{}
\shortauthors{Berentzen, Shlosman and Jogee}

\def\gtorder{\mathrel{\raise.3ex\hbox{$>$}\mkern-14mu
    \lower0.6ex\hbox{$\sim$}}}
\def\ltorder{\mathrel{\raise.3ex\hbox{$<$}\mkern-14mu
    \lower0.6ex\hbox{$\sim$}}}

\newcommand{\Bl}{ \left( }
\newcommand{\Br}{ \right)}

\begin{document}

\title{Stellar Bar Evolution in Cuspy and Flat-Cored Triaxial
       CDM Halos}

\author{Ingo Berentzen and Isaac Shlosman}

\affil{
Department of Physics and Astronomy,
University of Kentucky,
Lexington, KY 40506-0055,
USA \\ 
email: {\tt iberent@pa.uky.edu, shlosman@pa.uky.edu}
}
\and
\author{Shardha Jogee}
\affil{
Department of Astronomy, University of Texas at Austin,
1~University Station C1400, Austin, TX 78712, USA \\
email: {\tt sj@astro.as.utexas.edu}
}

\begin{abstract}
We analyze the  formation and evolution of stellar bars in galactic disks
embedded in {\it mildly} triaxial cold dark matter (CDM) halos that have
density
distributions ranging from large flat cores to cuspy profiles. We have 
applied tailored numerical simulations of analytical and live halos which 
include the feedback from disk/bar system onto the halo in order to test  
and extend earlier work by El-Zant \& Shlosman (2002). The latter employed 
the method of Liapunov exponents to analyze the fate of bars in analytical 
asymmetric halos. We find the following: (1) The bar growth is very similar 
in all {\it rigid} axisymmetric and triaxial halos. (2) Bars in live models 
experience vertical buckling instability and the formation of a pseudo-bulge 
with a boxy/peanut shape, while bars in rigid halos do not buckle. (3) In 
{\it live} axisymmetric halos, the bar strength varies by a factor of 
$\ltorder 2$, in growth or decay, during the secular evolution following the
buckling. The bar pattern speed evolution (i.e., deceleration) anticorrelates
with the halo core size. In such halos, the bar strength is 
larger for smaller disk-to-halo mass ratios D/H within disk radii, the bar 
size correlates with the halo core sizes, and the bar pattern speeds 
correlate with the halo central mass concentration. In contrast, bars 
embedded in {\it live} triaxial halos have a starkly different fate: they 
dissolve on a timescale of $\sim 1.5-5$~Gyr  due to the onset of chaos over 
continuous zones, sometimes leaving behind a weak oval distortion. The onset 
of chaos is related to the halo triaxiality, the fast rotating bar and the 
halo cuspiness. Before the bar dissolves, the region outside it develops
strong spiral structures, especially in the live triaxial halos. (4) More 
angular momentum is absorbed (fractionally) by the triaxial halos as 
compared to the axisymmetric models. The disk--halo angular momentum exchange
is mediated by the lower resonances in the latter models. (5) Cuspy halos are
more susceptible 
than flat-core halos to having their prolateness washed out by the action of 
the bar. The subsequent evolution is then similar to the case of a cuspy 
axisymmetric halos. We analyze the above results on disk and bar 
evolution in terms of the stability of trajectories and development of chaos 
in the system. We set important constraints on the triaxiality of DM halos 
by comparing our predictions to recent observational results on the 
properties of bars out to intermediate redshifts $z\sim 1$.   
\end{abstract}

\keywords{galaxies: evolution -- galaxies: formation -- galaxies:
halos -- galaxies: kinematics and dynamics -- galaxies: structure --
cosmology: dark matter}


\section{Introduction}

Stellar bars are recognized as the single most important
{\it internal} factor which drives the evolution of disk galaxies both
dynamically and secularly, modifying their morphology in this process. Modern
understanding of the bar growth is based on the efficiency of angular
momentum exchange between the (bar-forming) inner disk and the
surrounding dark matter halo, outer disk, bulge, and, to a certain
degree, with the immediate galactic environment (Athanassoula 2003). A
number of yet to be identified and investigated intricacies can affect
the efficiency of this process. 

In this work we attempt to analyze the formation and evolution of stellar 
bars embedded in fully grown disks in the presence of 
{\it triaxial}\,\footnote{We also use the term `asymmetric' concurrently with
`triaxial'}  
halos with various radial density profiles, from cuspy to flat-core models.
We study both analytical (i.e., rigid) and live triaxial halos and compare our
results with the evolution of the bars in axisymmetric ones. Corollaries for
disk evolution, such as the back-reaction of disks and bars on the halo
shapes, are investigated as well. This effect has broad implications for the
cosmological evolution of galaxies. Additional evolutionary effects on the 3-D
structure of stellar bars embedded in such halos will be addressed elsewhere.  

The evolution of bars is expected to be substantially altered if the
surrounding halos are even mildly non-axisymmetric (El-Zant \& Shlosman 2002,
hereafter ES02). However, this effect in a live disk-halo system
was not verified so far. Competing gravitational torques from a bar and a halo
acting on the main families of planar and 3-D periodic orbits can destabilize
them and dramatically reduce their ability to trap neighboring
trajectories, thus inducing chaos and dissolving otherwise stable structural
features in disk galaxies. The full extent of these non-linear effects is not
yet clear, but one can expect them to speed up secular changes in the 
collisionless components and to facilitate the angular momentum loss in the 
gaseous one. Although triaxial shapes of dark matter halos appear inherent in 
the cosmological numerical simulations (e.g., Barnes \& Efstathiou 1987;
Frenk et al. 1988; Dubinski \& Carlberg 1991; Warren et al. 1992; Cole \& Lacey
1996), relatively little attention has been paid so far to this issue when building
self-consistent galactic models. 

Observational constraints on the {\it shapes} of galactic halos coming from 
gas kinematics in disk galaxies (e.g., Sparke 1986;
Sackett 1999; Andersen et al. 2001; Merrifield 2002; Dekel \& Shlosman 1983)
and 
their polar rings (Sackett et al. 1994; Sackett \& Pogge 1995), gravitational
lensing (Kochanek 1995; Oguri et al. 2003; Hoekstra, Yee \& Gladders 2004) and
X-ray gas in ellipticals (Buote \& Canizares 1994, 1996; Buote et al. 2002)
remain inconclusive. 

Nevertheless, some clues exist. Residual {\it potential} axial ratios (both
flatness and prolateness\footnote{We define the halo flatness as $f=1-c/a$ and
its prolateness (i.e., equatorial ellipticity) as $\epsilon_{\rm H}=1-b/a$})
of about $0.9$ in Cold Dark Matter (hereafter CDM) halos are plausible,
even in {\it present} day galaxies (e.g., Kuijken \& Tremaine 1994 [for the
Milky Way]; Rix \& Zaritsky 1995; Helmi 2004). Another evidence for
triaxiality of the DM halo is provided by the X-ray isophote position angle
twist observed by {\it Chandra} in NGC~720 (but not in stellar
isophotes!) (Buote et al. 2002). A nearly prolate DM halo, with $T\sim 0.985$
(and $b/a\sim 0.791,$ $c/a\sim 0.787$) has been inferred for the
elliptical galaxy NGC~5128 from a kinematic study of planetary nebula system 
(Peng et al. 2004). We conjecture, therefore, that while some individual halo
shapes are compatible with being mildly prolate even at the present time, the
statistical significance of this effect is not clear, and it is not clear
whether disk galaxies differ in this respect from ellipticals. Overall, the
prolateness of {\it contemporary} halos appears to be insignificant. 

The main motivation behind this work is that theoretically a galactic halo is 
expected to acquire its triaxial shape during its initial collapse and to
support this shape during the ongoing process of a hierarchical merging. The 
degree of triaxiality\footnote{The triaxiality is defined here as
$T=[1-(b/a)^2]/[1-(c/a)^2]$, where $c/a$ is halo's polar-to-longest equatorial
axis ratio and $b/a$ -- equatorial axis ratio. $T=1$ 
corresponds to a prolate halo, while $T=0$ to an oblate one} will depend on the 
merging history, specifically on the relative angular momenta of the merger
precursors and interaction frequency --- which is difficult to quantify. 

Numerical simulations indicate that at very high redshifts, at the epoch of
galaxy formation, the halos can be significantly triaxial.
Detailed properties of individual numerical halos, such as triaxial
shapes and radial density structure, can be extracted from the models and
directly confronted by their observational counterparts. Flatness and
prolateness in the models appear to increase with the halo mass, being
somewhat milder in the $\Lambda$CDM than CDM cosmology
(e.g., for a recent review Bullock 2002). This triaxiality seems to
be marginally higher in the outer halo parts, independently of the
halo mass and of the ratio of rotational-to-dispersion velocity,
indicating that the halos are supported by anisotropic velocity
dispersions. 

The high halo triaxiality that may be present at the epoch of early galaxy
formation can be substantially diluted by the present-day. For instance, 
the addition of spherical and/or axisymmetric baryonic components
to the system (e.g., during the formation and development of a galactic disk)
can wash out the halo triaxiality, still keeping it non-negligible (Dubinski 
1994; Kazantzidis et al. 2004).  
 
From an observational standpoint it appears that {\it profiles} of
the DM halos in fully formed galaxies tend to have nearly constant density 
cores (Flores \& Primack 1994; Moore 1994; Burkert 1995; Kravtsov et al. 1998;
Borriello \& Salucci 2001; de Blok \& Bosma 2002; Gentile et al. 2004; etc.). A sim
ilar
effect has been recently observed in galactic clusters (Sand et al. 2002).
At the same time, dissipationless CDM simulations of galactic halos
agree with a universal density profile $\propto r^{-\alpha}$, where
$\alpha = 1 - 1.5$ (e.g., Dubinski \& Carlberg 1991; Warren et al. 1992; 
Crone et al. 1994; Cole \& Lacey 1996; Tormen et al. 1997;
Huss et al. 1999; Fukushige \& Makino 1997; Moore et al. 1999; Klypin et al.
2000; Jing \& Suto 2002; Power et al. 2003). (Note, Jing
\& Suto (2000) claim that the profiles are not
universal.) Navarro, Frenk \& White (1996, hereafter NFW) have found
a fitting formula for the density profile of DM halos, for a wide
range of cosmological  models, in which the inner profile diverges as
$r^{-1}$, while the outer profile drops as $r^{-3}$. It has been also
demonstrated theoretically that a cuspy density profile arises
inherently from the cold gravitational collapse in an expanding
universe (Lokas \& Hoffman 2000). The CDM model, therefore, predicts that the
inner density profile of galactic scale DM halos is characterized by
a density cusp while observations of the dynamics of the central
regions of galaxies imply a core-halo structure of the DM. Another
disagreement with observations is the so-called angular momentum
catastrophe --- the $N$-body and gas dynamical simulations
consistently result in too small galactic disks due to the overall
lack of angular momentum necessary to reproduce the observed disk
sizes (e.g., Burkert \& D'Onglia 2004).

This controversy between observations of DM cores and density cusps
in numerical models is not a fundamental one and can be resolved
within the general context of CDM cosmology. Within the conventional
physics framework, interactions with the dissipative and clumpy
baryonic component, such as dynamical friction, during the initial
stages of collapse can level off the central density cusps (which have
been shown to be thermodynamically improbable) and produce harmonic
cores in DM halos (El-Zant, Shlosman \& Hoffman 2001). Even though this may
affect the isodensity contours, making them rounder, it need not symmetrize the
isopotentials; these can remain asymmetric if triaxiality is not
affected beyond some radius (for example, a uniform bar has a
non-axisymmetric force contribution  inside its density figure).
More generally, this process was shown to replace the DM cusps with
baryonic cusps  (El-Zant et al. 2004). Furthermore, asymmetric and flat core
halos can have interesting implications for the disk growth and correlate
the properties of the central supermassive black holes with those of galactic
bulges and halos themselves (El-Zant et al. 2003). Alternatively, the central
density cusps have been proposed to dissolve by the action of galactic bars
(Weinberg \& Katz 2002; but see McMillan \& Dehnen 2005). 

A number of approaches can be taken in order to construct triaxial 
halos and elliptical galaxies (e.g., Aarseth \& Binney 1974; Merritt \& Valluri
1996; Holley-Bockelmann et al. 2001; Moore et al. 2004) and to study the bar
dynamics within triaxial halos. Ideta \& Hozumi (2000) have used analytical
density distributions for two highly prolate halos with (equatorial) axial
ratios of 0.6 and 0.75 and a very steep density distribution outside the core.
Here we choose to construct stable triaxial models with a subsequent
introduction of axisymmetric collisionless disks which are then released and
their time evolution is observed. In our initial conditions we tend to follow
ES02, who studied systematically the stability of stellar
bars in {\it analytical} halos of different triaxiality and central
concentration by means of Liapunov exponents. While in ES02 approach the
feedback from disk/bar system onto the halo is naturally ignored,
our models presented here fully account for it.
 
In ES02, mildly triaxial shapes have been used, with the gravitational
potential axial ratios in the equatorial plane, $b/a$, between 0.9 and 1.0,
and $c/a=0.8$, where $c$ is the halo polar axis. A clear trend has been
found, in the sense of models becoming intrinsically chaotic with
growing triaxiality and central concentration. For small halo (flat) core
sizes, $\sim 0.5 - 2$~kpc, and potential axis ratios of
$0.9-0.95$ most of the trajectories integrated appeared chaotic and
had large Liapunov exponents. Importantly, trapping by neighboring stability
islands is insignificant, because the distribution of values of Lyapunov
exponents is very similar to the distribution of occupied configuration
space volume. This means that stellar bars under
these conditions would disintegrate on timescales of a few dynamical
times, much shorter than the Hubble time, as chaotic trajectories
quickly diffuse out of the bar's configuration space. Even
spherically-symmetrical models, with a small core size, showed a
healthy fraction of chaotic trajectories, though connected regions of
regular orbits aligned with the bar remained in this case and a
self-consistent bar could be maintained.

What are the reasons for a dramatic increase in the fraction of chaotic 
orbits in the barred disks with the increase in central concentration 
(i.e., cuspiness) and triaxiality in the halo models of ES02? 
First, in terms of invariants of motion,
a stable 3-D figure must be built from trajectories which at least 
approximately conserve them. The chaos appears when the number of invariants 
of motion becomes smaller that the dimensionality of the system.
While in the flat core systems, the potential can be approximated
as quadratic (i.e., harmonic) and motions along the coordinates are independent
of each other, in cuspy potentials these motions are coupled, which leads to
the overall decrease in the number of invariants of motion and typically to a
chaotic behavior. In other words, cuspy density distribution cause solutions
for the Poisson equation to be far from quadratic, i.e., when the potential is
expanded in power series, it will have a non-negligible contribution from
terms beyond the quadratic one. These terms produce the coupling
between different degrees of freedom in equations of motion which become
highly non-linear. Such systems are prime candidates to excite chaotic motions
when perturbed.

The second reason is related to the time-dependent character of the
azimuthal force field comprised from the fast rotating\footnote{By `fast
rotating bar' we mean a bar which extends to about its corotation radius} 
bar and a non- or slowly tumbling halo. In other words, the origin of chaos in
this case lies in the comparable (in value) and competing forces from
the bar and the asymmetric halo. (Note, that in a halo potential only, most of 
the orbits are regular, even in the triaxial case.)

This onset of chaos in the presence of a bar within even a mildly triaxial halo
hints that such configurations are structurally unstable on dynamical
timescales. Since numerical cosmological halos are both triaxial and
centrally-concentrated, serious questions arise about the survival of
large-scale stellar bars or of the halos's triaxiality under these conditions.
This issue defines the general thrust of our work.

This paper is structured as follows. In the next section we provide the initial
conditions for numerical modeling. Sections~3 and 4 describe our results for
analytical and live models, and section~5 is devoted to discussion and
concluding remarks.

\section{Numerical Modeling}

\begin{figure*}[ht]
\begin{center}
\includegraphics[angle=-90,scale=0.70]{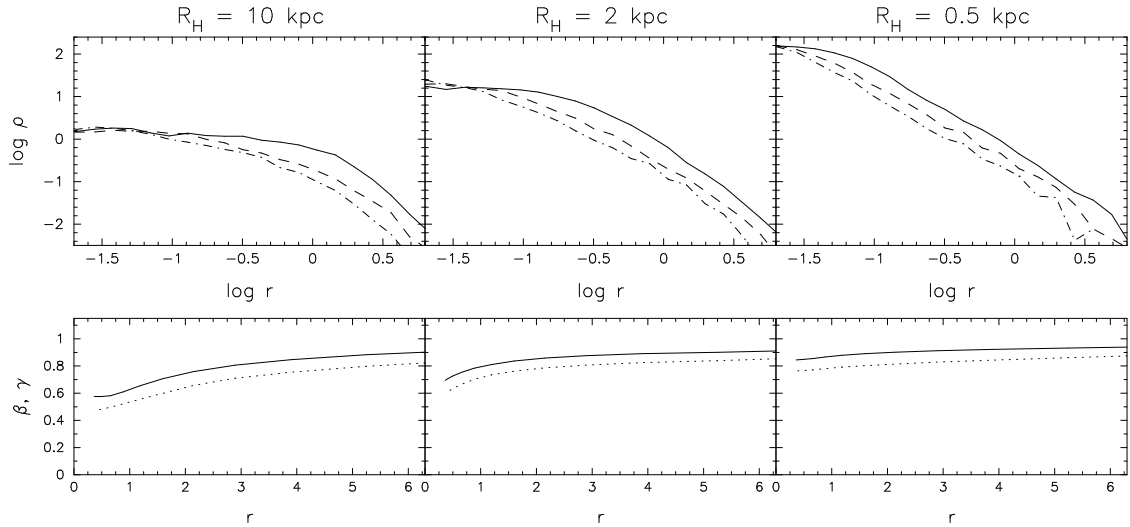}
\end{center}
\caption{{\it Top panels:} Density distribution of the {\it live} triaxial
halos along its principal axes (solid line: along $a$, dashed line: along $b$,
dotted line: along $c$-axis) for models with a varying core size, $R_{\rm H}$.
{\it Bottom panels:} Potential axes ratios $\beta = b/a$ (solid line) and
$\gamma= c/a$ (dotted line) for the same models.}
\label{plot1}
\end{figure*}

We have introduced the following dimensionless model units. The spatial
distance unit is taken as $r\!=\!10$\,kpc, the mass unit is
$M\!=\!10^{11}~{\rm M_\odot}$ and the gravitational constant is chosen
to be ${\rm G}\!=\!1$, which result in a time unit of
$\tau\!=(r^3/{\rm G}\,M)^{1/2}=4.7\times 10^7$\,yrs, corresponding
to the dynamical timescale, $\tau_{\rm dyn}$. In these units, the velocity
is given in $208~{\rm km~s^{-1}}$. The actual physical units
are used when it is needed for clarity.

We have used version FTM-4.4 of $N$-body code (e.g., Heller \& Shlosman 1994;
Heller 1995) with $N\!=\!6-9\times 10^5$ particles and gravitational softening
of 100~pc to simulate the
collisionless disk and spheroidal galactic components (i.e., bulges and DM
halos) in a large number of
models. Our results appear to be reasonably independent of $N$. The
gravitational forces have been computed using Dehnen's (2002) {\tt falcON}
force solver, a tree code with mutual cell-cell interactions and complexity
{\it O(N)}. It conserves momentum exactly and is about 10 times faster than
optimally coded Barnes \& Hut (1986) tree code.
 
To analyze the angular momentum redistribution in the models, we have applied
the spectral analysis method described in Binney \& Spergel (1982), in 
conjunction with our nonlinear orbit finding algorithm (e.g., Heller \&
Shlosman 1996). Athanassoula (2002) has shown that the angular momentum
exchange between the disk and the halo is mediated by the lower
resonances, with the disk losing and the halo gaining the angular momentum.
Overall, the resonances provide an independent test of the quality of
the numerical scheme. When the $N$-body potential is too `grainy', the
particles cannot be locked by the resonances and so the latter efficiency
in redistributing the angular momentum is sharply reduced.   

\subsection{Initial conditions}

\begin{figure*}[ht]
\begin{center}
\includegraphics[angle=-90,scale=0.5]{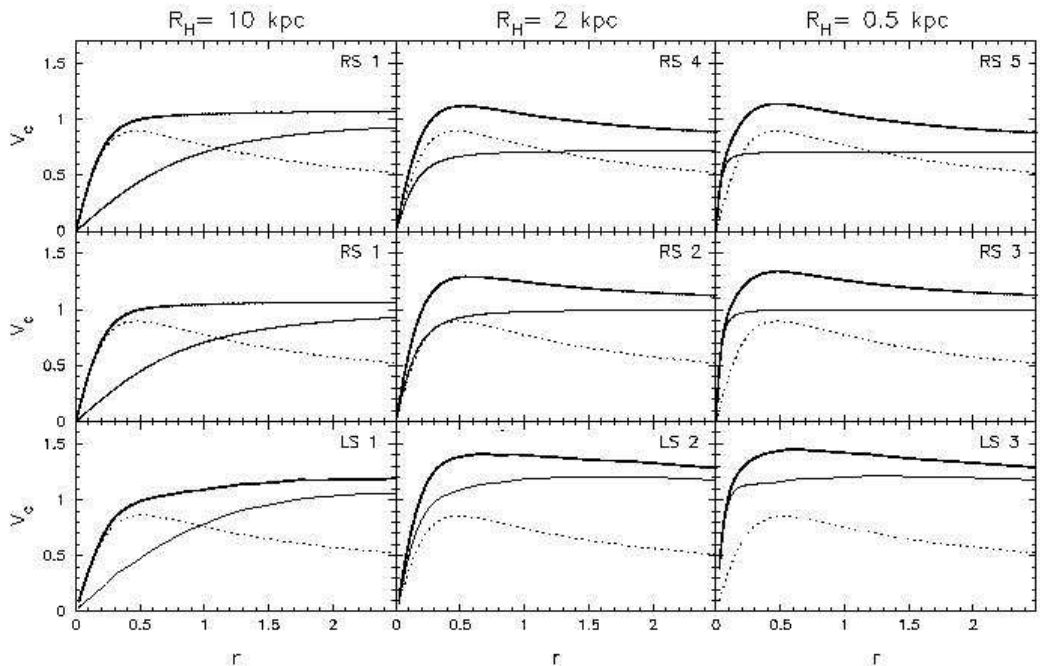}
\end{center}
\caption{Initial rotation curves for ({\it upper panel:}) RS\,1,
RS\,4 and RS\,5 models (rigid spherically-symmetric halos, constant $M_{\rm D}/M_{\rm
H}$) within $r=10$~kpc;
({\it middle panel:}) RS\,1 to RS\,3 models (rigid spherically-symmetric halos,
constant $V_{\rm H}$); and ({\it lower panel:}) LS\,1 to LS\,3 (live
axisymmetric halos) with cores of 10, 2 and 0.5\,kpc. Thin solid line shows
halo contribution, dotted line --- disk contribution, and thick solid line
--- the total. The units of length and velocity correspond to our model
units.}
\label{vcurve}
\end{figure*}

\begin{deluxetable*}{lcccclccl}
\tablecaption{INITIAL MODEL PARAMETERS}
\tablehead{
Model & $R_{\rm H} [kpc]$ & $V_{\rm H}$ & $\beta$ & $\gamma$ & Notes }
\startdata
{\bf RS\,1} & 10.0 & 1.0 & 1.0 & 1.0 & Rigid spherically-symmetric halo \\
{\bf RS\,2} &  2.0 & 1.0 & 1.0 & 1.0 & Rigid spherically-symmetric halo \\
{\bf RS\,3} &  0.5 & 1.0 & 1.0 & 1.0 & Rigid spherically-symmetric halo \\
\hline

{\bf RS\,4} &  2.0 & 0.721 & 1.0 & 1.0 & Rigid spherically-symmetric halo \\
{\bf RS\,5} &  0.5 & 0.708 & 1.0 & 1.0 & Rigid spherically-symmetric halo \\
\hline

{\bf LS\,1} & 10.1 & 1.13 & 1.0 & 0.99 & Live axisymmetric halo \\
{\bf LS\,2} &  1.8 & 1.18 & 1.0 & 0.98 & Live axisymmetric halo \\
{\bf LS\,3} &  0.3 & 1.20 & 1.0 & 0.98 & Live axisymmetric halo \\ 
\hline

{\bf RT\,1} & 10.0 & 1.0 & 0.9 & 0.9 & Rigid triaxial halo \\
{\bf RT\,2} &  2.0 & 1.0 & 0.9 & 0.9 & Rigid triaxial halo \\
{\bf RT\,3} &  0.5 & 1.0 & 0.9 & 0.9 & Rigid triaxial halo \\ \hline

{\bf RT\,4} &  2.0 & 0.719 & 0.9 & 0.9 & Rigid triaxial halo \\
{\bf RT\,5} &  0.5 & 0.708 & 0.9 & 0.9 & Rigid triaxial halo \\ 
\hline

{\bf LT\,1} & 10.5 & 1.14 & 0.79 & 0.72 & Live triaxial halo \\
{\bf LT\,2} &  2.4 & 1.19 & 0.84 & 0.77 & Live triaxial halo \\
{\bf LT\,3} &  0.5 & 1.19 & 0.88 & 0.79 & Live triaxial halo \\ \hline

{\bf LT\,3\,HM} &  0.7 & 1.10 & 0.84 & 0.77 & as LT\,3, but with half-mass disk
\\
\hline

{\bf LT\,3\,MT} &  0.5 & 0.4 & 0.9 & 0.8 & as LT\,3, but maintaining
triaxiality
\enddata
\label{table:models}
\tablecomments{Columns: (1) model type (see text); (2) (fitted) halo core
radius; (3) (fitted) constant in eq.~1; (4) (fitted) halo equatorial axial
ratios; (5) (fitted) halo polar-to-longest axis ratios}
\end{deluxetable*}

The main challenge in running evolutionary models of galactic disks
embedded in triaxial halos is to form the initially stable halo
configurations in the first place. A limited number of options known
to us was not considered to be satisfactory, as they did not allow for
a sufficient control of initial triaxiality and mass distribution in the
halos. Instead we have designed the following procedure to obtain the 
required halo properties, which is described below.

First, we have used the known density-potential pairs of required
symmetry to lay out particles up to some given radial cut-off radius.
The velocity distribution function has been taken $\propto E^{-\alpha}$,
where $E$ is the energy and $\alpha = 2.5-3.5$, i.e., below the $3.5$
value for the Plummer sphere. Next, the halo has been evolved for
about $50\,\tau_{\rm dyn}$ to allow for any residual relaxation to
cease. This has resulted in a stable spherically-symmetric
configuration. For the axisymmetric halos, a {\em frozen} stellar
disk has been gradually introduced thereafter and the halo was been
given time to relax again in this growing disk potential. The disk
potential slightly changes the initial halo core size and therefore
we apply an adiabatic cooling procedure to the halo particles within
the core region, by gradually reducing the velocities.
Before introducing the disk for the triaxial cases, we have implemented
a similar adiabatic `heating-cooling' procedure to the halo particles
to create a triaxial halo of required flatness and prolateness, by
using `heating' of the velocities along the $x$-axis and `cooling' along both
the $y$- and the $z$-axes, over some period of time, i.e.  some $40\,\tau_{\rm
dyn}$. The fractional energy heating/cooling per dynamical time is very small,
$\sim 10^{-6}$ per machine timestep (corresponding to about $10^{-5}$ per
dynamical time), so the system is never taken out of a dynamical equilibrium.
The disk potential partially dilutes the halo triaxiality and therefore
we apply a second heating-cooling procedure after the disc has been
introduced in the models. This finally results
in both the required $\beta\!=\!b/a$ and $\gamma\!=\!c/a$ distributions
and core sizes in the halo potential (Fig.~\ref{plot1}).
Note that in the models with smaller cores it is more difficult to
impose (and maintain) the triaxiality at all radii.

For a direct comparison with ES02 in this work, we aimed to
closely reproduce their initial conditions. A 3-D mass distribution (live or
analytical) which pairs with the nonrotating logarithmic
potential, $\Phi_{\rm H}$, (e.g., Binney \& Tremaine 1987) was used
for the halo, 
 
\begin{equation}
\Phi_{\rm H} = \frac{1}{2}
\,V^{2}_{\rm H}\, \log \left(R_{\rm H}^{2} +
x^{2} + \beta^{-2} y^{2} + \gamma^{-2} z^2 \right)   \ ,  
\label{logarithm}
\end{equation}

\noindent 
where $V_{\rm H}$ is the asymptotic (in the limit $R \gg R_{\rm H}$
and $\beta, \gamma \rightarrow 1$) circular velocity, $R_{\rm H}$ is
the halo core radius and $\beta~(=b/a)$, $\gamma~(=c/a)< 1$ are the
azimuthal and polar potential axis ratios.
Note that the corresponding mass distribution is diverging for
this potential. We, therefore, applied a truncation radius
for models with live halos, i.e. $50$ and $100$\,kpc for the
axisymmetric and triaxial halos, respectively.
Although the potential of these models is going to change, both as
a result of the disk addition and, in the case of live halo response,
to the disk evolution, we find it nevertheless beneficial to compare
the evolving potential to the fitted logarithmic one. For spherical
(axisymmetric) models $\beta=\gamma=1.0$, while for triaxial models
$\beta\!=0.9$ and $\gamma\!=0.8$, approximately for live models, after
the halo has relaxed in the disk potential.  

We have used two sets of initial halo models. In the {\it first set},
$1\rightarrow 4\rightarrow 5$ (see Table~1) ---
for rigid halos only -- we adjust $V_{\rm H}$ requiring $1\!:\!1$
ratio of halo-to-disk mass within $r\!=\!1$. This means that on the
scale of $r\sim 1$ models with different core size have the same mass
concentration. On smaller scales of course the more cuspy models will
be always more centrally concentrated, which can affect the properties
of developing stellar bars.

In the {\it second set}, $1\rightarrow 2\rightarrow 3$, for which we 
run both rigid and live models,
we retain the value of $V_{\rm H}$, obtained for an axisymmetric
halo with a core radius of $R_{\rm H}\!=\!1$. For further halo
configurations, the total halo mass within its truncation radius
$r\!=\!100$\,kpc, is $10.0$ in our units. Thus, when moving from a
large flat core to cuspy models, we move from a maximal disk model
to a halo-dominated one.

Similarly, as in ES02, the stellar disk has been set up following the potential
in the form of (Miyamoto \& Nagai 1975),
 
\begin{equation}
\Phi_{\rm D} = -  \frac{{\rm G}\,M_{\rm D}} {\sqrt{ x^{2}+y^{2}+ \Bl A_{\rm D}
+\sqrt{B_{\rm D}^{2}+z^{2}} \Br^{2}
} } \ ,    
\label{miyamoto}
\end{equation}
 
\noindent which describes a disk-bulge system, with the
parameters $A_{\rm D}\!=\!0.284$ and $B_{\rm D}\!=0.05$, determining the
scalelength and height, respectively. The disk mass is $0.5$ within $r\!=\!1$ 
and about $0.6$ within the disk cut-off radius, i.e. $r\!=\!2.5$.
The initial Toomre's $Q$ parameter is taken $1.2$ and constant with
radius. The disk rotational velocities in triaxial halos have been set using 
our standard procedure but ignoring the mild triaxialities in the total 
potential. The initial models are summarized in Table~\ref{table:models} and 
the initial rotation curves are shown in Fig.~\ref{vcurve}. 

\newpage
\section{Results}

We now present the bar evolution and its interaction with the outer disk and
halo for the numerical models listed in Table~\ref{table:models}. We start
with disks in rigid axisymmetric\footnote{We are interested only in this
property of spherical halos and refer to them as axisymmetric}  halos and
compare them with those in
rigid triaxial halos, assuming constant disk-to-halo mass ratios within
10~kpc, and varying the flat core sizes from large 10~kpc to cuspy ones. 
We then continue with live axisymmetric and triaxial halos. 

\begin{figure*}[ht]
\begin{center}
\includegraphics[angle=-90,scale=0.75]{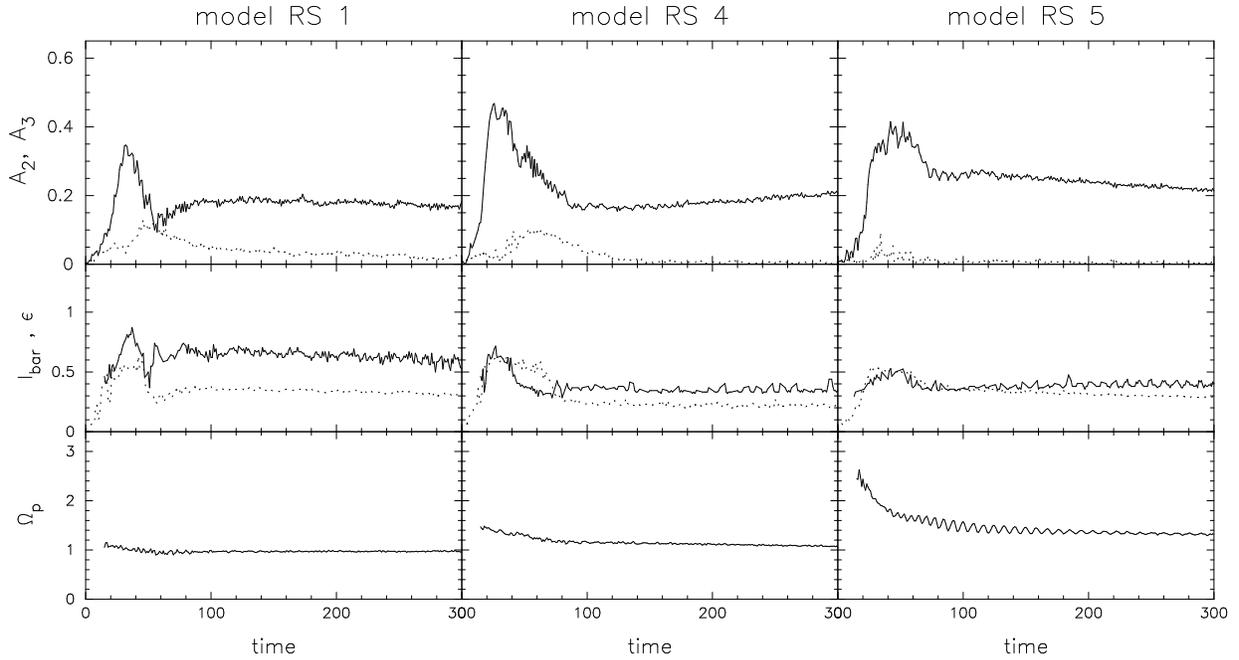}
\end{center}
\caption{Bar strength (top panels; $m=2$ amplitude $A_2$: solid 
line, $m=3$
amplitude $A_3$: dotted line), bar length and (max.) ellipticity (middle
panels; solid and dotted line, respectively) and pattern speed $\Omega_{\rm
p}$ (bottom panels) as a function of time for models~RS\,1, 4 and 5
(from left to right) wit rigid axisymmetric halos.}  
\label{plot2a}
\end{figure*}

\begin{figure*}[ht]
\begin{center}
\includegraphics[angle=-90,scale=0.75]{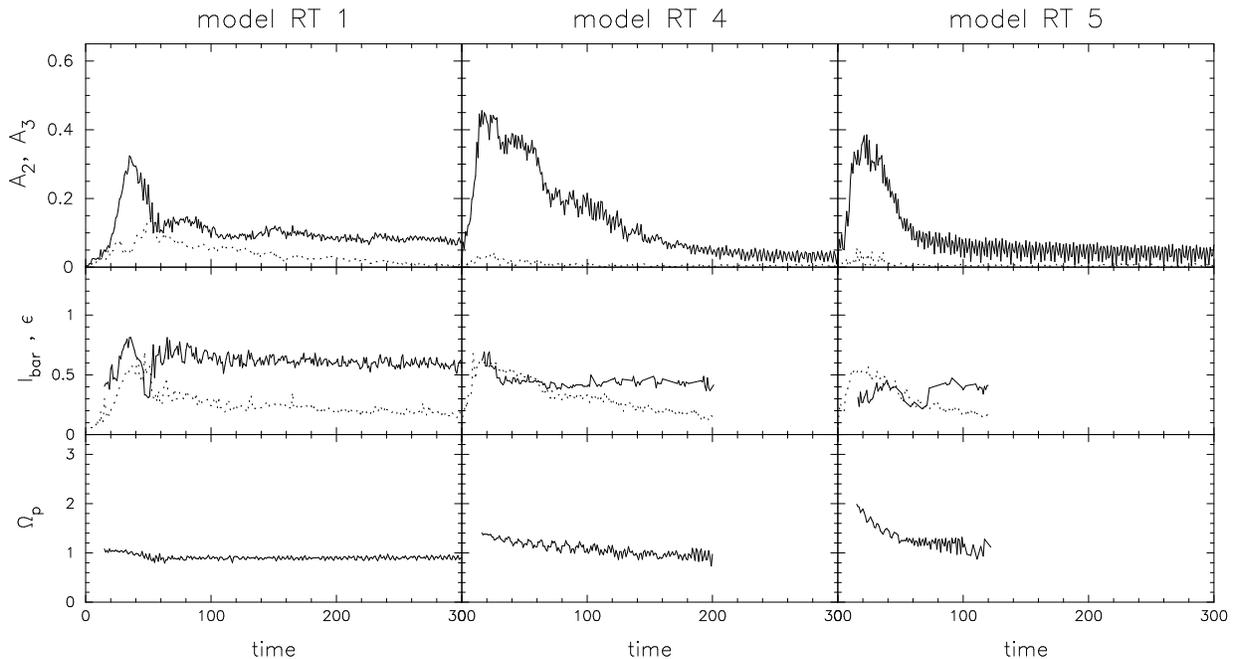}
\end{center}
\caption{Same as Fig.~\ref{plot2a}, but for models with rigid triaxial
  halos, i.e. RT\,1, 4 and 5 -- constant mass ratio within 10~kpc.}
\label{plot4a}
\end{figure*}

\subsection{Comparing Bars in Rigid Axisymmetric and Triaxial Halos with 
Constant Disk-to-Halo Mass Ratios}

\subsubsection{Rigid axisymmetric halos}

Only circumstantial evidence exists comparing the bar formation and
evolution in analytical (i.e., rigid) and live halos, even in the case of an
axial symmetry (e.g., Christodoulou, Shlosman \& Tohline 1995 and refs.
therein), with the exception of Athanassoula (2002). 
It is known that rigid halos are less hospitable to bar
growth. All equal, bar instability in this case requires larger disk-to-halo
mass ratio than in the live halo systems. This behavior of stellar disks
embedded in non-responsive halo potentials can be readily understood in terms
of absence of (resonance) halo orbits capable of absorbing the disk angular
momentum. As expected, it is accompanied by a nearly constant bar pattern
speed, as the angular momentum loss from the bar is minimized to the outer
disk only.

Fig.~\ref{plot2a} exhibits some properties of stellar bars in such rigid
axisymmetric (RS) halos --- with a large flat-core $R_{\rm H}\!=\!10.0$\,kpc
(model~RS\,1), a smaller core $R_{\rm H}=2.0$\,kpc (model~RS\,4)
and a cuspy (model~RS\,5; $R_{\rm H}=0.5$\,kpc) density profile
(see Table~\ref{table:models}). The (normalized) amplitude of the $m\!=\!2$
mode, $A_2$ (Fig.~\ref{plot2a}, top panels) provides only a partial description
of the bar evolution. Because it is an integrated quantity over some specific
radius and because the bar sizes will differ from model to model,
it does not allow for a direct comparison between bars in different models.
One can encounter a small size but strong bar whose amplitude will be
diluted (say) by a large averaging region. To resolve this dilemma, we use
both the bar amplitude, measured within a thin cylinder of radius of 5\,kpc,
and the bar length and its maximum ellipticity, $\epsilon = 1\!-b/a$ (obtained
from the isodensity fits, in the plane
(Fig.~\ref{plot2a}, middle panels), to characterize its strength. Both
approaches can be tested observationally.

We note that a reliable determination of a $N$-body bar size from isophote
fitting has its pitfalls (see Martinez-Valpuesta \& Shlosman 2004). Before the
vertical buckling, the bars have a flat distribution of ellipticity with
radius, while at later times the ellipticity has a clear maximum. We find
empirically
that a good approximation to the bar size would be the radius where the bar
ellipticity drops by 0.1. This is verified using nonlinear orbit analysis,
finding the $x$-extent of the largest stable $x_1$ orbit in the bar 
(Martinez-Valpuesta, Shlosman \& Heller 2006). On the
other hand, the ellipticity of the surrounding disk shows a gradual decline
with radius, even in mildly triaxial halos used in this work. This different
behavior of ellipticity in numerical bars and disks allows us to safely
separate them.

The rigid axisymmetric halo models RS\,1, 4 and 5 have been arranged
along the same sequence of decreasing core size as in Fig.~\ref{vcurve}.
These models have the same disk-to-halo mass ratio within the central
10~kpc, but correspond to increased central concentration in the halo. 

All three models develop a strong bar in the process of a `normal' bar
instability (Fig.~3). The bar amplitudes, $A_2$, show a strong peak
after the initial bar growth, at $\tau\sim 30-50$, and a subsequent decline,
partly associated with a transient $m=3$ mode $A_3$. The latter appears to be 
a purely numerical artifact resulting from mixing of analytical (halo) and 
live (disk) potentials. The more centrally-concentrated models supress the
(planar!) $m\!=\!3$ mode more efficiently. The sudden weakening and shortening
of the bars has been related to the onset of chaos in strong bars
(Martinez-Valpuesta \& Shlosman 2004), which can be the case in these models
as well. Asymptotically, the more cuspy models show marginally
stronger bars. While the RS4 bar exhibits a slight secular growth in $A_2$, 
the RS\,5 bar shows a secular decline at the same time, again probably related
to stronger chaos excited in cuspy models.

The asymptotic bar size, $\sim 6-7$~kpc, is largest in RS\,1 --- the large halo
core model, and about $3-4$~kpc in RS5. The {\it
maximal} bar size (at $\tau\sim 30-50$) also clearly
correlates with the halo core size, $R_{\rm H}$. This trend is much
more pronounced in models which have increasing central mass concentration
(see section 3.2.1 and Fig.~5).The bar ellipticity, $\epsilon$, closely
follows the size evolution of the bars in all models. The bar pattern speed,
$\Omega_{\rm p}$, shows some initial decline with time, more substantial in
cuspy models, but then levels off --- as expected, because of lack of angular
momentum absorbing material in axisymmetric and to a certain degree also in
mildly triaxial halos. The outer disk quickly saturates for this
redistribution of angular momentum. As expected, more cuspy
models have progressively faster $\Omega_{\rm p}$.

We find that the bars in rigid halos are not subject to the buckling
instability (e.g., Pfenniger \& Friedli 1991), apparently because of
difficulty to develop vertical asymmetric modes, $m=1$. But vertical $m=3$ and
5 show up progressively more in RS\,4 and RS\,5, at the level of $\sim 8\%$. 
The vertical resonant heating in the bar (e.g., Friedli \& Pfenniger 1990) is
not obviously observed, possibly due to the rigid potential of the halo,
and especially due to the absence of a discreteness noise in the analytical
potential. 

The effect of the bar on the disk is to shorten the original radial
scalelength in the inner (bar region) disk and in the outer disk. In this
sense, the combination of initial conditions and disk evolution maintain a
double-exponential disk, $r_{\rm
D}$, characterized by the inner and outer radial scalelengths. The inner
disk (the bar region) in RS\,1 develops $r_{\rm D}\sim 2.2$~kpc, when measured
along the bar major axis. This value is progressively smaller for more cuspy
models, e.g., $\sim 1.6$~kpc in RS\,5. In the outer disk,  $r_{\rm D}$
decreases from its initial value $5.7$\,kpc to about 5\,kpc with time and
remains constant when disk/bar reaches quasistationary stage after $\tau\sim
80$ (in RS\,1).
The Toomre's $Q$ increases overall from 1.2 to 2.1, measured at two radial
initial scalelengths after the bar weakening in RS1 and more so in RS\,4 and
RS\,5.
 
The vertical disk scaleheight, $z_{\rm D}$, shows progressively less change
with increased cuspiness. While it increases from 0.5\,kpc to $\sim 0.7$~kpc
in RS\,1, as a result of disk heating, towards the end of the simulations, it
increases only 0.6~kpc in RS\,4 and stays unchanged in RS\,5. 
 
\subsubsection{Rigid triaxial halos}

In this section we describe the evolution of the models with rigid
triaxial halos and a constant $M_{\rm D}/M_{\rm H}$ within $10$\,kpc
(Table~\ref{table:models}), i.e. models RT\,1,  RT\,4
and RT\,5. Some of the bar properties in these models are shown
in Fig.~\ref{plot4a}. The angular momentum redistribution is discussed
in section~4.

The bar initial growth and size, its maximal strength (i.e., of $A_2$), and 
even its peak ellipticity in rigid triaxial halos are remarkably 
similar to those in axisymmetric ones. The only difference is that the growth 
rate is faster and so the peak strength is achieved earlier. The subsequent
evolution, however, is very different --- the bar dissolves over the 
characteristic time of $\sim
30-100$, depending on the model, leaving only a weak oval distortion behind
(Fig.~4). The general trend of bar evolution in this sequence of models
clearly links the halo cuspiness to the dissolution process, which is
amplified by the halo triaxiality --- a trend predicted by ES02 based
on the orbital stability analysis and the development of connected
chaotic domains in these models

The bar size plot (mid panels), which indicates a bar of $6$\,kpc in RT\,1 at
later times, is misleading here and seems to represent only the disk response
to the triaxial potential and the ellipticity is that of the disk. As  long as
the bars exist in these models, their pattern speeds mirror those in
axisymmetric halos of the same mass concentration.

Despite the destruction of bars in RT\,1 to RT\,5 models, the disk tends to
acquire
a double-valued radial exponential scalelength: the inner one of about 2~kpc
and the outer one of $\sim 5-6$~kpc. The Toomre's $Q$ and the vertical
scaleheight behave as in the axisymmetric sequence. 
 
\subsection{Comparing Bars in Rigid and Live Axisymmetric Halos with Decreasing
Disk-to-Halo Mass Ratios}

\subsubsection{Rigid axisymmetric halos}

\begin{figure*}[ht]
\begin{center}
\includegraphics[angle=-90,scale=0.75]{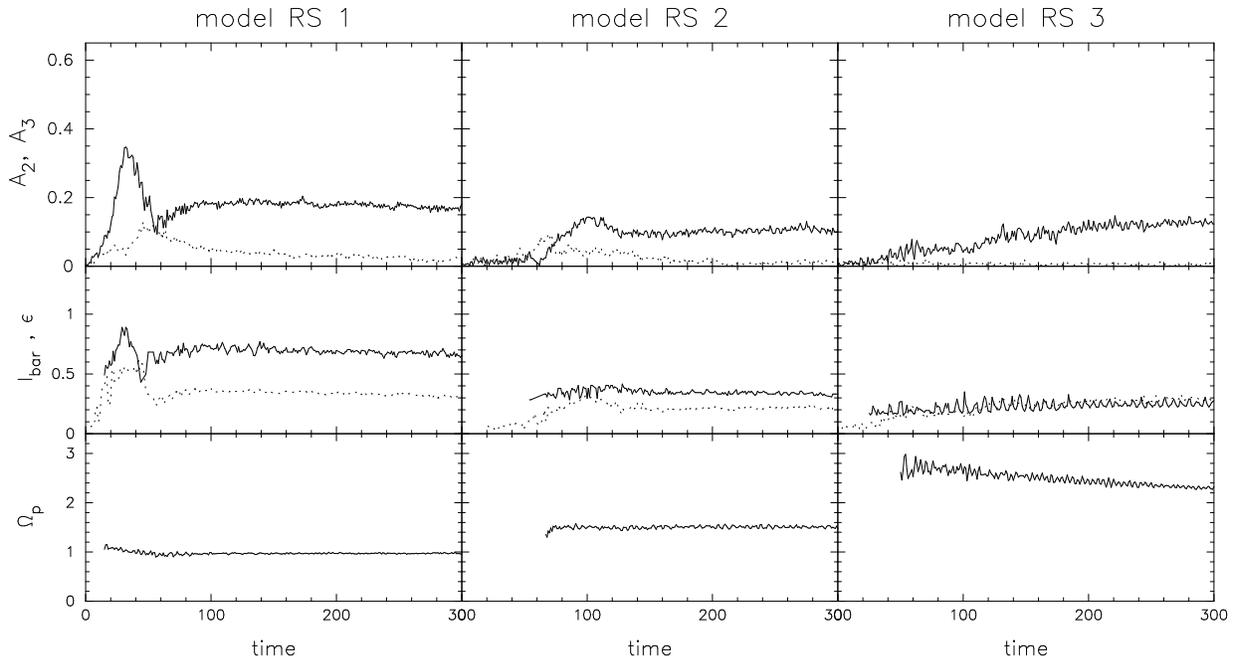}
\end{center}
\caption{Same as Fig.~\ref{plot2a}, but for models with rigid axisymmetric
halos, i.e. RS\,1, 2 and 3 -- from disk to halo dominated.
}
\label{plot2}
\end{figure*}

To provide a more direct comparison to the results of ES02, we present
in this section a set of simulations in which we closely follow their
model parameters. The basic difference with the previous set of models is
that halo mass within the central 10~kpc is increased for progressively
smaller cores (Table~1). Thus, when moving from flat core to cuspy models, 
we move from a maximal disk model (RS\,1) to a halo-dominated one (RS\,3) 
as evident from Fig.~2. Fig.~\ref{plot2} shows the evolution of stellar bars 
in such rigid axisymmetric halos. This corresponds to the increased central 
concentration, in excess of that shown by 
RS\,1$\rightarrow$RS\,4$\rightarrow$RS\,5 sequence. 

Fig.~5 exhibits the disk evolution inside such halos. The strong bar only
develops in the model RS1. The RS2 supresses the bar growth for the first
$\tau\sim 60$, developing a short, $\sim 2-3$~kpc bar. Thereafter, $A_2$
drops and the bar further weakens to a rather oval distortion which
persists over the period of simulations. The RS3 supresses the bar growth rate
even stronger, and $A_2$ shows a steady secular growth not leveling off even 
after $\tau\sim 300$, while $A_3$ is completely supressed in a more cuspy 
models.  

$Q$ increases to $\sim 2$ while the vertical scalelength shows no change
whatsoever. The bar pattern speed correlates with the halo's mass
concentration, i.e. increasing with smaller core sizes. Pattern
speed in RS\,3 is linearly decreasing with time reflecting the secular 
increase in $A_2$.
 
To summarize the sequence of rigid axisymmetric halos RS\,1$\rightarrow$RS\,3:
the stellar bar is basically supressed in more cuspy models in accordance
with ES02 analysis.  

\subsubsection{Live axisymmetric halos}

\begin{figure*}[t]
\begin{center}
\includegraphics[angle=-90,scale=0.75]{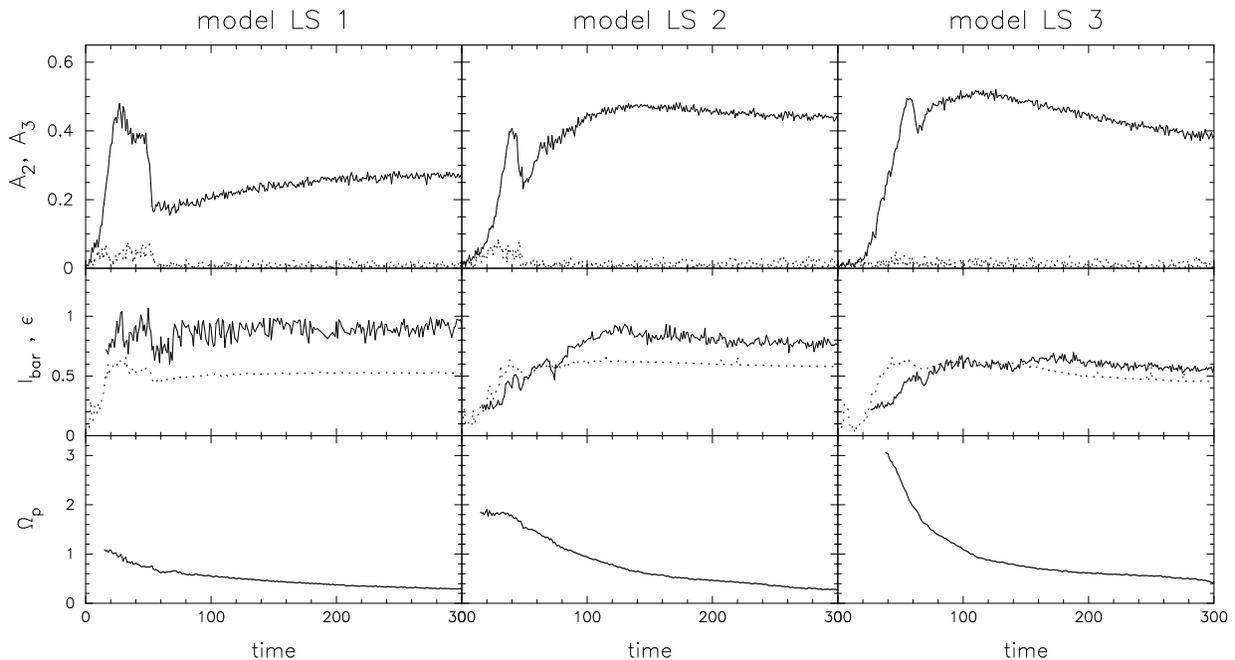}
\end{center}
\caption{Same as Fig.\ref{plot2a}, but for models LS\,1--3 with live
 axisymmetric halos.}
\label{plot3}
\end{figure*}

\begin{figure}[t]
\begin{center}
\includegraphics[angle=-90,scale=0.90]{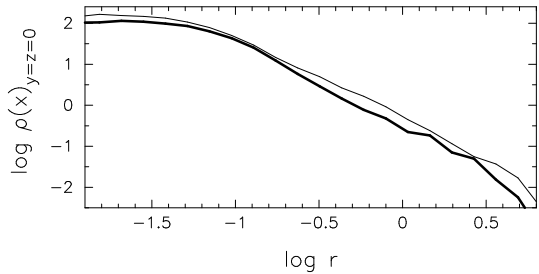}
\end{center}
\caption{Density profile along the halo main (longest) axis for
 time $t=0$ (thin line) and $t=300$ (thick line) for model LS\,3
 ($r_{\rm H} = 0.05$).}
\label{plot12}
\end{figure}

Models LS\,1 to LS\,3 present the case of live axisymmetric and
increasingly cuspy halos. After introducing the disk and letting the halo to
relax, we have fitted the halo density profile at the end
of this relaxation process using the parameters of logarithmic
potential, $V_{\rm H}$, $R_{\rm H}$, $\beta$ and $\gamma$
(see Table~\ref{table:models}). The new density
profiles for the halo can be fit reasonably well and show slight increase in
$V_{\rm H}$ and unchanged $\beta$. Owing to the flattened disk potential
$\gamma$ slightly decreases. For model LS\,1, the core radius
remained at $\sim 10$~kpc, for model LS\,2 it decreased to $\sim 1.8$~kpc,
and for model LS\,3 --- to some $0.3$~kpc. The main difference between
these models and the previously discussed ones with rigid halos,
however, is an active redistribution of angular momentum between the
bar forming region in the disk and the live halo (see section~4). 
For all live axisymmetric halos during disk evolution
the halo density profile is stable (e.g., as shown for LS\,3 in Fig.~7).

The sequence LS\,1$\rightarrow $LS\,3 has produced the strongest bars among
all of our models (Fig.~6). The amplitude $A_2$ has also shown a pattern of
behavior:
(1) the bar growth is fastest in LS\,1 (disk dominated) and slowest in LS\,3 
(halo dominated); (2) the maximal bar strength is increasing along the
sequence; (3) all bars exhibit vertical buckling, (4) $A_2$ decays most
strongly in LS1 after the buckling, and least in LS\,3; and finally (5) the
post-buckling bar in LS\,1 continuous its secular growth, which quickly
saturates in LS2 and starts to decay in LS\,3. Hence LS\,1 (and to a lesser
degree
LS\,2) shows most resemblance to the secular evolution of a stellar bar in
Martinez-Valpuesta, Shlosman \& Heller (2006) where the bar weakens
dramatically during its buckling and increases its strength thereafter. 

For model~LS\,1, the bar size decreases from roughly $10$\,kpc just before
the buckling, to $\sim 7$\,kpc at $\tau\!\sim\!60$ and then increases again to
about $9$\,kpc after the buckling. For model~LS\,3, it grows from
$5$\,kpc to $8$\,kpc. This means that during the initial growth,
up to the time of the buckling, the bar is confined to the regular
region delineated in ES02.

The bar sizes show clear correlation with the halo core size, $R_{\rm H}$,
as in the rigid sequence RS1 $\rightarrow$ RS5 described in section~3.1.
One can clearly observe here the `delay' in the bar growth in
halo-dominated models. The bar ellipticity shows a visible decay after
the buckling in LS1 and stabilizes afterwards, while in LS\,2 and LS\,3
$\epsilon$ does not decay.  

As expected, the bars slow down substantially over the simulation time and
the initial slowdown is well correlated with $R_{\rm H}$
(Fig.~\ref{plot3}, bottom panels). This angular momentum transfer from the disk
is deposited mostly in the internal circulation in the halo as its figure does
not acquire rotation (section~4).

We have mentioned above that all LS models show buckling which is less
pronounced for halo-dominated models. This can be noted from the amplitudes of
the vertical modes ---  LS\,1 shows clear increase of $m\!=\!1$, and LS\,2 and
3 only in $m\!=\!3$ and $5$. 

The disk evolution in LS models again leads to a double-exponential
scalelength: the inner bar-dominated part has $r_{\rm D}$ which slightly
decreases from LS1 to LS3, from $\sim 2$~kpc to $\sim 1.6$~kpc, while the
outer disk scalelength actually increases along the sequence, from about 5~kpc
to just below 6~kpc. 

The disk scaleheight in LS1 shows an abrupt increase at $\tau\!\sim \!50$ from
its initial value 0.5 to about 0.9\,kpc within few dynamical times, during the
vertical buckling of the bar. This effect is clearly related to the formation
of a pseudo-bulge, i.e., bulge of a boxy/peanut shape. LS\,2 experiences a
milder 
increase in $z_{\rm D}$ to
$\sim 0.65$~kpc and in LS\,3 it remains largely unchanged.
 
All LS models show the formation of pseudo-bulges associated with the vertical
buckling of the bars. However the shapes of these bulges differ and this
difference persists for the time of the simulations. While in model LS\,1 and
LS\,2 the isodensity contours show a more boxy shape, the
cuspy model, LS\,3 develops a peanut-shaped bulge (see also Athanassoula \&
Misiriotis 2002).
 
\subsection{Comparing Bars in Rigid and Live Triaxial Halos}

\subsubsection{Rigid triaxial halos}

\begin{figure*}[t]
\begin{center}
\includegraphics[angle=-90,scale=0.75]{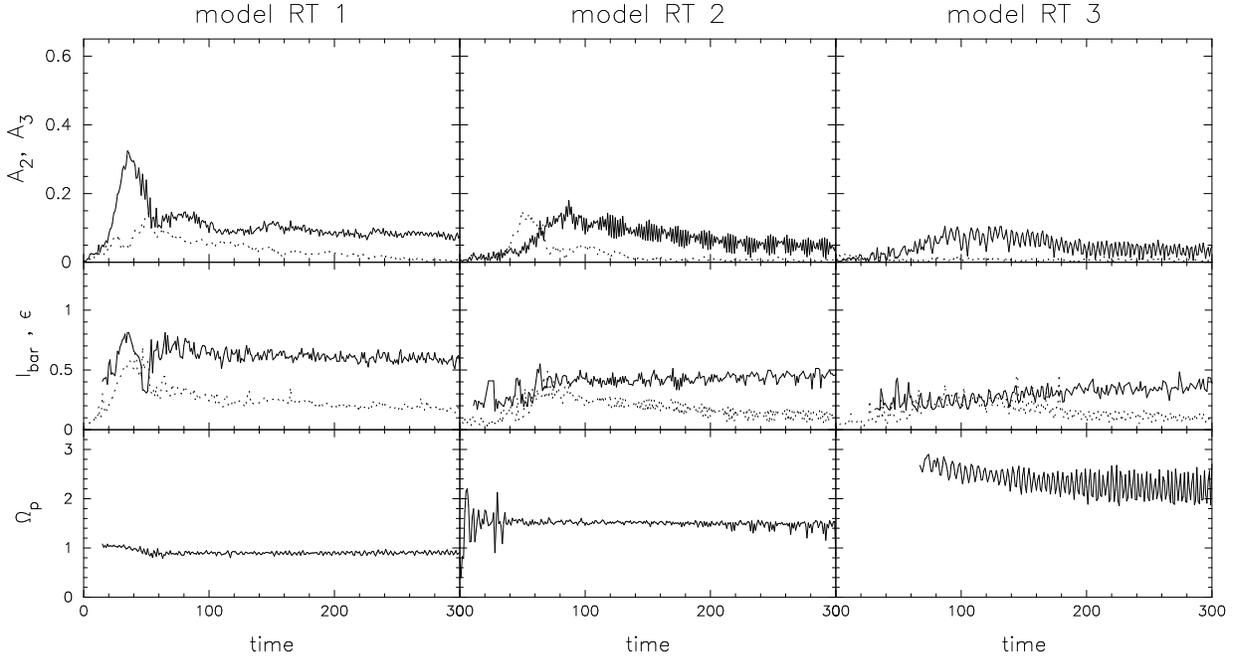}
\end{center}
\caption{Same as Fig.~\ref{plot2a}, but for models with rigid triaxial
halos.}
\label{plot4}
\end{figure*}

The RT\,1$\rightarrow$RT\,3 sequence describes the disk evolution within rigid
triaxial halos with decreasing core size, closely following the initial
conditions in ES02 (Fig.~8). The only difference between the previously
discussed RT1$\rightarrow$RT5 sequence and RT\,1--3 is that the latter one is
progressively halo-dominated. This difference is responsible for the damping
of bar instability in a disk embedded in RT\,2 and RT\,3 halos. Even RT\,1
shows
a substantial decay of the initially strong bar to a largely an oval
distortion, unlike the bar in RS\,1 model. Hence, asymptotically RT\,2 and
RT\,3
behave as RT\,4 and RT\,5. However, early in the disk evolution, the former
models show strong bars (see section~3.2.1). 

The size and ellipticity of the central (bar) oval distortion in RT\,1 to RT\,3
correlate well with $R_{\rm H}$. In RT\,1, the maximal bar length is $\sim
9$\,kpc at $\tau\sim 35$ and after $\tau\sim 50-60$, the bar deteriorates into
a rather triaxial (in the equatorial plane) configuration within the central
$r\sim 4-5$\,kpc. Note  the $A_2$ amplitude becomes finite already in the first
moments into the simulation because of the disk response to the halo
triaxiality. 

The first few disk rotations, especially in RT2 and RT\,3, show a grand design
spiral structure in the outer disk which remains tightly bound and quickly
decays. Even more than in RS\,3, the disk stays largely axisymmetric
within the central few kpc after $\tau\sim 30$. Its surface density
profile shows (almost) no evolution. The outer disk radial scalelength
decreases to about 4.5--5\,kpc, while the disk scaleheight stays 
unchanged. 

\subsubsection{Live triaxial halos}

As in the case of the live axisymmetric halos, the triaxial halos are modified
by the introduction of the frozen disk potential. At the moment of disk
`release,' the halo core size in model LT\,1 has slightly increased to $R_{\rm
H}\!\sim 10.5$~kpc, in LT\,2 it has increased to $\sim 2.4$~kpc, and in
LT\,3 stayed at $0.5$~kpc (see Table~1). We found it increasingly difficult
to maintain the (inner) halo triaxiality in cuspy models after introduction
of a frozen disk. This has affected the LT\,3 model which possesses a more
structurally unstable halo. 

\begin{figure*}[ht]
\begin{center}
\includegraphics[angle=-90,scale=0.75]{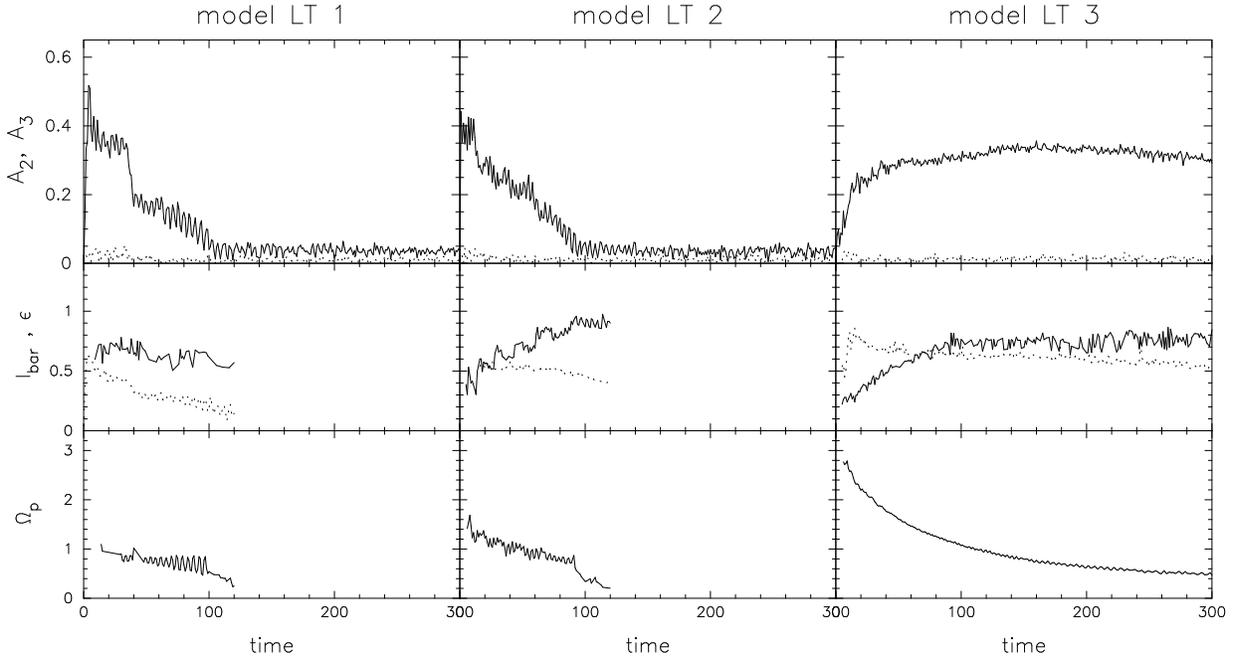}
\end{center}
\caption{Same as Fig.~\ref{plot2a}, but for models with live triaxial
halos.}
\label{plot5}
\end{figure*}

\begin{figure*}[ht]
\begin{center}
\includegraphics[angle=-90,scale=0.75]{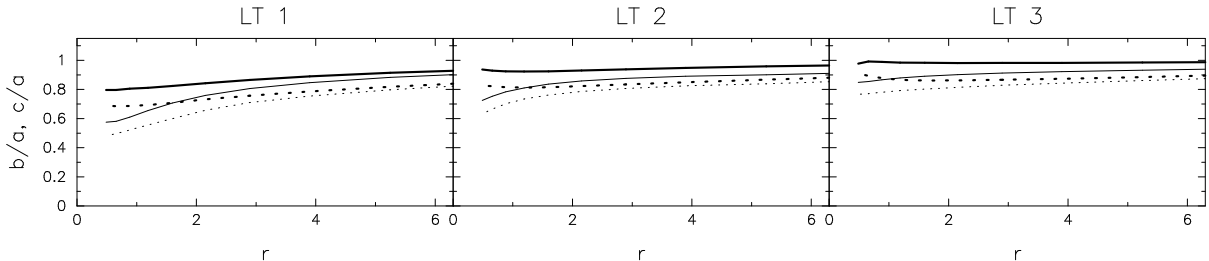}
\end{center}
\caption{Halo (isopotential) axes ratios $\beta$ (full lines) and $\gamma$
(dotted lines) as a function of radius for models with live triaxial
halos. Thin and thick lines indicate the profiles at the beginning
and the end ($t\!=300$) of the runs, respectively.}
\label{plot10}
\end{figure*}

The evolution of $A_2$ amplitude in the LT\,1$\rightarrow$LT\,3 sequence is
remarkably different from that in the rigid sequence RT\,1 to RT\,3.
The initial disk response to the halo potential has induced strong
bars in all these models. The fate of the bar, however, differs
depending on the model. The bars decay nearly linearly in LT\,1 and LT\,2
on a timescale of $\tau \sim 100$, and enter a steady state in LT\,3.
Overall, the LT\,3 model behaves differently and its evolution
resembes more the LS\,3 model, although the bar is not so strong. The
buckling is clearly visible in all LT models. However, the resulting boxy
shapes of bulges are quickly erased in LT\,1 and LT\,2, while peanut shape
bulge in LT\,3 persists till the end of the simulation. The
planar $A_3$ amplitude is negligible in all live halo models.

The radial density profile in the triaxial halo is stable during live disk
evolution in all these models, even for the cuspy LT3. However the halo
triaxiality $T$ is reduced sharply in LT\,3, due to decrease in $\epsilon_{\rm
H}$.
After $\tau\sim 70$, it has largely lost its prolateness, $\beta\sim 1$
(Fig.~10). This halo axisymmetrization is only partially a result of initial
conditions --- LT\,3 has had somewhat larger $\beta$ at $\tau=0$, as noted
above.
Rather, as we discuss in section~5, the halo is much more structurally
unstable in cuspy models. In order to test this latter assumption, we have
re-run the LT2 model with a {\it frozen} disk. Fig.~11 shows the resulting
evolution of potential axial ratios for the halo. Both halo flatness, $f$, and
its prolateness, $\epsilon_{\rm H}$, show no evolution over 300 dynamical
times, $\sim 14$~Gyrs. This confirms that it is the developing bar
perturbation which decreases the halo prolateness without much effect on its
flatness.  

\begin{figure}[ht]
\begin{center}
\includegraphics[angle=-90,scale=0.85]{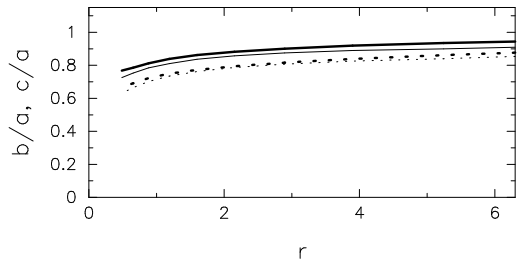}
\end{center}
\caption{Testing the halo evolution in the LT2 model: re-run of the model with
a {\it frozen} instead of the live disk, for comparison. Axes-ratio $\beta$
(full line) and $\gamma$ (dotted line). The thin and thick lines represent the
times $t=0$ and $t=300$, respectively. The absence of evolution in $\beta$ and
$\gamma$ in this case confirms that the evolution observed in Fig.~10 results
from interaction between the disk and the halo.}
\label{plot11}
\end{figure}

The bar size and ellipticity LT\,1 and LT\,2 (Fig.~9) represent a combined disk
and bar evolution. The outer disk develops elaborate system of long-lived
spiral arms, much more pronounced than in live axisymmetric models. Because
of the difficulty to disentangle between the bar, spiral arms and the oval
disk, the estimated bar size in Fig.~9 is erroneous. The
inner disk radial scalelength shows a general decrease to about 1.7~kpc in
LT\,1 and to $\sim 1.8$~kpc in LT\,2 and LT\,3. The outer disk has $r_{\rm D}$
approaching $\sim 4.8$~kpc in LT\,1 and about $5-5.5$~kpc in LT\,2 and LT\,3.
The inner disk scaleheight increases abruptly to $\sim 0.9$~kpc after 
$\tau\!\sim \!35$ in LT\,1 and to $\sim 0.8$~kpc in LT\,2 and $\sim 0.75$~kpc
LT\,3 models.

The live triaxial models are especially efficient in triggering the spiral
structure in the barred disk. The arms penetrate deeper towards the center
in LT\,2 model with smaller core radius. The strength of the spirals depends on
the position angle of the bar with respect to the longest halo axis --- when
the bar is normal to this axis, the spirals are prominent. This type of spiral
regeneration happens as long as the bar exists.   

\subsubsection{Models with continuous support for halo triaxiality}

Models LT\,1 to LT\,3 show that the disk (and bar) response to a live triaxial
(halo) potential can act as to decrease its prolateness, e.g., equatorial
ellipticity. This effect is most pronounced in model LT\,3, where it has been
difficult to induce the halo's triaxiality, especially in its central region. 
Compared to the models with larger cores, the disk in LT\,3 evolves more
similarly to that in axisymmetric halos. Because of the strong feedback of
stellar bars onto the halo triaxiality, we have run a model
for LT\,3 with a continuous support for triaxiality --- the LT\,3\,MT model. 
The simple justification for this is that mergers will induce triaxiality in
the extended DM halos in the first place. We, therefore, mimic this process by
maintaining or continuously regenerating a mild halo triaxiality, and at the
same time allowing the live halo to interact with the live disk. 

\begin{figure*}[ht!!!!!!!!!!!!!!!!!!!!!!!!!!!!!!!!!!!!!!!]
\begin{center}
\includegraphics[angle=-90,scale=0.75]{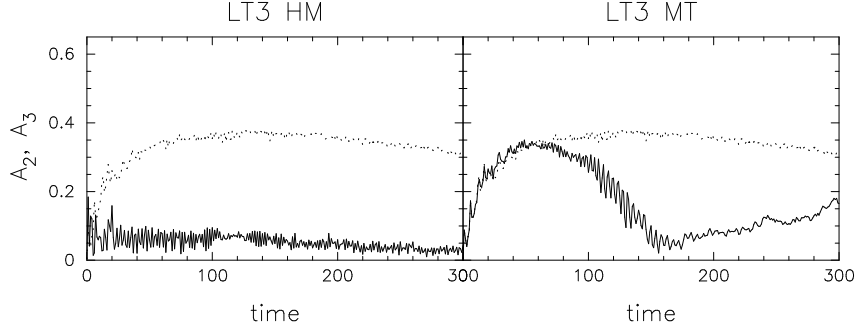}
\end{center}
\caption{The bar amplitude $A_2$ for models LT\,3\,HM (0.5~kpc core) with a
half mass disk (left panel)
and LT\,3\,MT continuous support of triaxiality (right panel). The dotted
line shows $A_2$  of LT\,3 for comparison.}
\label{plot8}
\end{figure*}

Fig.~12 (right) exhibits such an evolution of LT\,3\,MT and provides $A_2$ from
LT\,3 for a comparison. The adiabatic heating/cooling procedure was gradually 
turned on
and has its full strength from $\tau = 70$ to 225, when it is gradually turned
off. Initially, the LT\,3\,MT evolved identically to LT\,3. Because the halo
triaxiality has been maintained at the steady level
(Table~1), it has a profound effect on the disk. The bar completely dissolves
and starts to grow afterwards when the triaxiality is gradually washed out. 
 
\subsubsection{Light disks in triaxial halos}

The interaction between the halo and the evolving disk tends to gradually wash
out the initial triaxiality in the halo. The efficiency of this 
process depends on the disk-to-halo mass ratio. The bar is then able to
grow within the region in which the halo becomes more axisymmetric. To show
explicitly the dependency of halo and disk evolution on their (inner) mass
ratios, we have run live triaxial models with lower disk-to-halo mass
ratio within central 10~kpc for the cuspy LT\,3 model. Here we show only
LT\,3\,HM --- the model with a half disk mass. The bar
instability in this model is substantially suppressed compared to LT\,3 model. 
LT\,3\,HM develops a weak oval distortion in the center
(Fig.~12, left) whose shape depends strongly on its orientation with respect to
the halo major axis --- its ellipticity is smaller when it is normal to the
halo, and larger, when it is parallel. Otherwise, the disk is stable and no bar
instability develops.

\section{Disk-Halo Angular Momentum Exchange}

\begin{figure*}[ht!!!!!!!!!!!!!!!!!!!!!!!!]
\begin{center}
\includegraphics[angle=-90,scale=0.75]{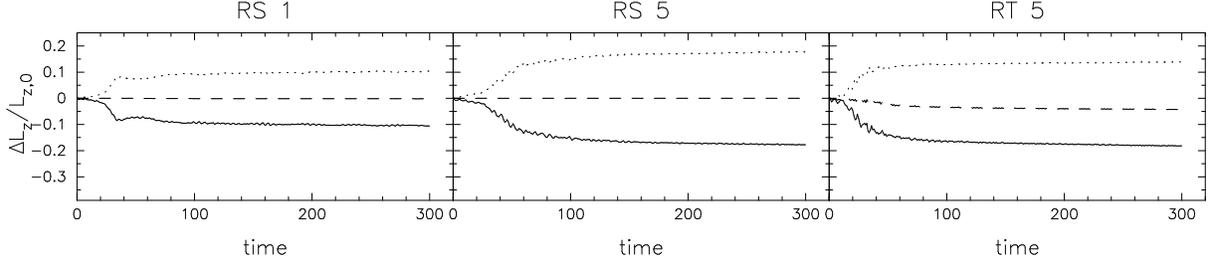}
\end{center}
\caption{Models RS\,1, RS\,5 and RT\,5 showing the angular momentum exchange in
the live disk along the sequence with increasing cuspiness and with a rigid
axisymmetric (RS\,1, RS\,5) and triaxial (RT\,5) halos. The change in the
angular momentum in the disk, $\Delta L_z$, is normalized by the total angular
momentum in the disk, $L_{z,0}$, at $\tau=0$ when it is released.  The solid
line represents integrated $\Delta L_z/L_{z,0}$ in the disk between
$r=0-10$~kpc, the dotted line --- for $r > 10$~kpc and the dashed line --- the
change in the total angular momentum in the disk, which is absorbed by the
halo.}
\label{plot8}
\end{figure*}
\begin{figure*}[ht!!!!!!!!!]
\begin{center}
\includegraphics[angle=-90,scale=0.75]{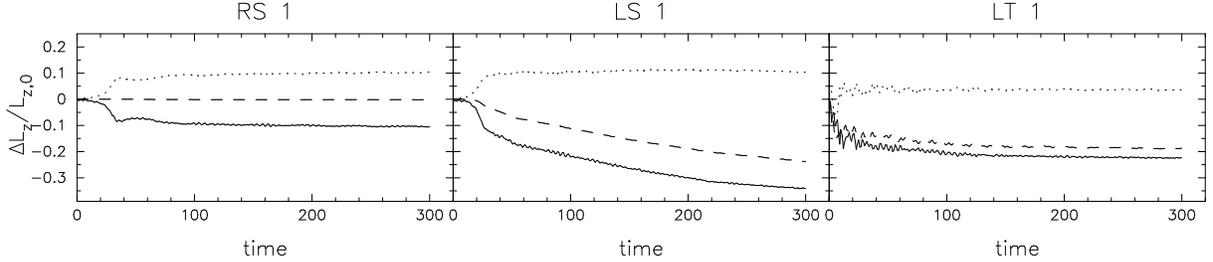}
\end{center}
\caption{Models RS\,1, LS\,1 and LT\,1 showing the angular momentum exchange in
a rigid vs live axisymmetric vs live triaxial halos. As in Fig.~13.}
\label{plot8}
\end{figure*}

The general direction of the angular momentum flow in all models is
expected to be from the inner bar-forming disk towards the outer disk and
the halo. This reflects behavior of all
models that break their axial symmetry for whatever reason, that of initial
conditions (triaxial halos) or as a result of a spontaneous bar instability.
However, when this redistribution is mediated by the bar, there is also an
additional flow from the inner disk (inside the bar corotation) to the outer
disk.   

In the models with rigid axisymmetric halos the exchange of angular momentum is
limited to that between the inner and outer disks, across the corotation and
mediated by the bar. Models RS\,1 and RS\,5 (Fig.~13) show that the total loss
of the angular momentum from the disk is $\Delta L_{\rm z} =0$ and that the
outer disk disk gains the angular momentum from inner part. This gain is
larger for RS\,5 as
it develops stronger bar. Rigid triaxial models drain some of the
angular momentum from the disk (e.g., RT\,5 in Fig.~13) but this does not
dominate the $L$-exchange because of the mild triaxiality here.  

Fig.~14 shows the effect of introduction of a live halo, axisymmetric and
triaxial. The total loss of the $L$ by the disk has been dramatically
amplified compared to the rigid triaxial halos. The outer disk still gains the 
same amount but most of $L$ goes to the halo. The strong and long-lived bar 
developing in the LS\,1 model is instrumental in this transfer.  
The main difference with the LT1 model is that the disk immediately
acquires a strongly oval shape which is gradually lost and so
overall the transfer amount of angular momentum is less than in LS\,1. However, 
during the
first 20--30 dynamical times, the $L$-transfer is stronger in LT\,1.
Subsequently, the bar which has developed in LT\,1 dissolves, which levels off
the rate of angular momentum trasfer.  

Furthermore, in order to test the effect of the resonance coupling between
the disk/bar and the halo, we have performed the spectral analysis of the
orbital motion (Binney \& Spergel 1982) in conjunction
with our nonlinear orbital finder algorithm (Heller \& Shlosman 1996).
We find that a fraction of disk particles in live axisymmetric halos is locked 
by the lower resonances,
especially by the inner Lindblad resonance (ILR). A fraction of the halo
particles is locked by the corotation resonance --- hence ILR-CR resonance
coupling. Following Athanassoula (2002), we have frozen the model potentials
at times $\tau=100$ and 150, and compared the angular
momentum of individual particles at these times. Most of the angular momentum
flow between the disk and the halo was mediated by this ILR-CR coupling.
Particles trapped at $\tau=100$ largely remained trapped at 150. This result
shows that the adverse effect of the grainy $N$-body potential on the evolution
in our models is limited, and the latter is driven by a chaotic dynamics
instead.

\section{Discussion and Conclusions}

We have analyzed the formation and evolution of stellar bars in galactic disks
embedded in rigid (i.e., analytical) and live axisymmetric and mildly-triaxial
DM halos of a varying cuspiness. The nearly axisymmetric potential employed
in our simulations allows us to focus on the development of bar instability
{\it per~se} and does not violate the overall dynamics and survival of the 
disks. Using tailored numerical simulations we aimed at testing and extending
the predictions
of El-Zant \& Shlosman (2002, ES02) which employed the method of Liapunov
exponents to address the fate of bars in analytical triaxial halos. Our
simulations of live halos, unlike ES02, include the feedback 
between the disk, bar, and halo. We fully support the main conclusions of ES02
and provide some additional insight into the bar and disk evolution. 

We start with summarizing our results and follow up with discussion --- 
\underline{\it first} based on the
bar strength. In all cases we can separate the initial bar
growth to its full strength, over $\tau\sim 30-50\sim
1.5-2$~Gyr, from the 
subsequent bar evolution, dynamical or secular, over the Hubble time. The first
stage is very similar between all {\it rigid} axisymmetric and triaxial halo 
models.  We find, that while some of the rigid halos suppress the bar
formation,
depending on the disk-to-halo mass ratio D/H within the disk radius, all live
halos develop impressive bars with ellipticities in the range of $\sim
0.4-0.7$, 
depending on their evolutionary stage. In axisymmetric live cases (only!) these 
bars appear stronger for smaller D/H, in line with Athanassoula \& Misiriotis 
(2002). In live halos, the bars develop faster in triaxial compared to 
axisymmetric halos. (This statement is marginally true also for rigid halos.) 

All bars weaken at the end of the first stage. This process is accompanied
by bar's vertical buckling in the live halos and formation of exponential
(pseudo-) bulges of a peanut or boxy shape. No buckling is observed in rigid
halos. 

The subsequent evolution of the bar differs substantially between 
axisymmetric and triaxial halos. In the former case, the bars appear
dynamically stable and show a limited secular evolution by a factor of
$\ltorder 
2$ in strength, in growth or decay after the initial vertical buckling. This 
result is in agreement with Athanassoula (2002). In the 
latter case, the bars dissolve almost completely on a timescale of $\tau\sim 
30-100 \sim 1.5-5$~Gyr, sometimes leaving behind a weak oval
distortion in the 
central regions. This is true for all models with the exception of LT\,3 --- 
a live cuspy triaxial halo, where the triaxiality (i.e. prolateness) is erased 
early due to the feedback from the bar and its ensuing evolution closely
resembles that of the axisymmetric halos. 

To verify that the bar endurance is related to the erasure of the halo 
prolateness, we have run a test model where (live) halo triaxiality in LT\,3
has been maintained at a constant level. The bar dissolved in this case 
as expected. On the other hand, to show that the evolution of halo prolateness
is due to destabilizing action of the bar, we run models with a frozen 
axisymmetric disk in live triaxial halos --- these models show basically no 
evolution for $\epsilon_{\rm H}$.  
We have also run a number of test models with the LT\,3 halo but 
smaller D/H mass ratio. Models with D/H$\ltorder 0.5$, formed only a weak oval 
distortion. This is unlike the trend in the axisymmetric live halos, where
small D/H resulted in stronger bars. We discuss this below. 

\underline{\it Second,} the bar sizes appear to correlate with the halo core
sizes, with the
exception of live triaxial models where the bar size is more difficult to
establish due to the strong response in the disk and a quick bar dissolution.
The bar ellipticities provide additional source of information on the bar 
strength and even more about the state of the disk when the bar amplitudes
measured by $A_2$ decay below 0.1. 

\underline{\it Third,} the pattern speeds of bars anticorrelate with halo core 
sizes --- more centrally concentrated halos produce faster tumbling bars. 
However, the deceleration rate of the bar anticorrelates with the halo core
sizes as well --- thus, after the Hubble time, the end pattern speed of the
bars is similar.
As expected, bars in rigid halos quickly reach constant pattern speeds, while
bars in live halos continue to slow down asymptotically. Interestingly, bars
in live axisymmetric halos can become stronger or weaker while slowing down.
Bars in live triaxial halos always weaken and dissolve while slowing down.

\underline{\it Fourth,} in rigid halos, the angular momentum of the inner disk 
is fed into the outer
disk, across the bar corotation. Some angular momentum is lost to the halo
in rigid triaxial models. In live models, the halo appears to gain angular
momentum, more so fractionally in triaxial models. We have verified that in
axisymmetric models the angular momentum
is mediated by lower resonances, especially by the inner Lindblad 
resonance in the disk and by the corotation resonance in the halo.

\underline{\it Finally,} in triaxial models, as long as a bar is present, 
the disk evolution is characterized by development of a strong spiral structure 
outside the bar region. It is especially prominent in live triaxial halos. 
The outer disk acquires exponential 
surface density distribution, while the inner (bar-unstable) part can be 
modeled roughly by a shorter exponential scalelength when measured along the 
bar major axis.    

Our emerging understanding of the evolution of bars embedded in live, mildly 
triaxial or axisymmetric halos is based on two physical processes which 
determine the fate of the system. 
These are the angular momentum redistribution in the disk-halo system and 
the development of chaos. We can make a general statement that angular momentum 
flows from the bar unstable region in the disk to the outer disk and the halo. 
This process is accompanied by the initial growth of the bar both in strength 
and in size. In axisymmetric halos, it ultimately saturates and the bar
enters a steady-state phase characterized by secular changes (Athanassoula 
2003). Instead, in a mildly triaxial halo which hosts a barred disk, the
dominant
process is the increase in the fraction of chaotic trajectories --- it
affects both the bar and halo structures. In other words, no steady-state
develops, and either the bar is dissolved or halo prolateness is washed out.
This effect on the bar embedded in a {\it rigid} triaxial halo has been 
calculated in ES02, and here we have presented fully self-consistent 
numerical simulations accounting for the bar and {\it live} halo evolution.

After the submission of this work we have learned about the work by Curir et al.
(2005) where stellar disks have been evolved within the cosmological DM halos,
albeit at much lower resolution than in our work. In these simulations, the
bars form in responce to the original torquing by the halo, but, contrary to
our main results, survive for the rest of the simulation. We strongly suspect
that this different behavior of bars results from the loss of prolateness
by DM halos in their models, similarly to our model LT\,3. Furthermore, it is 
the process of introduction of disk into the halo which is also responsible 
for washing out the halo prolateness (Section~2.1), which required us to
apply the second heating-cooling procedure. Unfortunately,
Curir et al. do not provide any analysis of the shape of their halos after bringing
up the disks and during their subsequent evolution. Their Table~1 refers to 
purely DM halos without stellar disks only.

Thus we expect that the combination of halo triaxiality and its cuspiness in
conjunction with a fast rotating\footnote{As defined in the footnote of
section~1} bar will lead to the development of chaos
and to the dissolution of the least massive or structurally stable triaxial
feature in the system, i.e., the bar or halo triaxiality. In section~1
(see also ES02 and refs. therein), we have argued that cuspy density 
distribution cause solutions for the Poisson equation to be far from quadratic
and this will produce the coupling between different degrees of freedom 
in equations of motion which become highly nonlinear --- a prime recipe to
develop a chaotic behavior when such a system is perturbed.   
 
To understand the results of these numerical simulations within the context
of developing chaos in a triaxial system with a fast tumbling stellar bar,
we take the next step and discuss the evolution of the bar strength in terms of
the
stability of trajectories used by ES02. Ultimately bars and other 
morphological features, such as prolate halos, dissolve when the density figure 
stops to support the figure of the background potential, in other words when 
self-consistency of density-potential pair is violated. As mentioned in
section~1,
a stable 3-D figure must be built from trajectories which at least 
approximately conserve the invariants of motion. Chaos appears when the number 
of invariants of motion becomes smaller that the dimensionality of the system.

Two issues appear most relevant in this respect: the overall asymmetry of the 
background potential and the degree of mass concentration. Stability of 
trajectories for potentials and density distributions
used in this work have been analyzed in ES02 by calculating the Liapunov 
exponents on a high-resolution 3-D cylindrical grid. These exponents provide a 
measure of timescales which are associated
with dynamical instabilities and are especially suitable to study development
of chaotic motions in time-dependent potentials (see detailed description
of this method in sections~1, 2 and the Appendix of ES02).   

The mapping of halos which are under consideration here using the 
Liapunov exponents has shown that triaxial halos are regular in the absence of
a bar --- the top panel of Fig.~6 in ES02 indicates that trajectories within
this potential are stable well over the Hubble time. This is
supported by live models of triaxial halos presented here. All the models
appear stable unless bar-unstable disks are added, {\it even the cuspy halo 
in} LT\,3, as expected. It is  the addition of a (triaxial) bar fast 
tumbling with respect to the 
halo, which destabilizes the trajectories and triggers the chaos. The middle 
panels in Fig.~6 of ES02 show this effect in a cuspy halo --- most of the 
configuration space becomes chaotic with the characteristic timescale 
$\ltorder 1$~Gyr. Moreover, the chaotic regions are fully connected, thus 
providing a strong indication that survival of any non-axisymmetric structure
is highly questionable. The less cuspy barred models of ES02 show progressively
less chaotic regions, especially in the flat cores, and the existing chaos is 
less interconnected. A self-consistent bar-disk model of Pfenniger (1984), 
without a surrounding halo, shows chaotic regions confined to the bar 
corotation radius (lower panels, Fig.~6 of ES02), in line with the above 
explanation. We also note that the remarkable correspondence between  the
distributions of the available configuration space volume for the trajectories
in ES02 and the maximal Liapunov exponents points out that the nearby stability
islands are not efficient in trapping the chaotic trajectories. Therefore,
one should expect that the predicted bar dissolution in ES02 models will
indeed be observed in numerical simulations.

The sequences, RT\,1$\rightarrow$RT\,3 and RT\,1$\rightarrow$RT\,5 discussed in
section~3 represent a gradual increase in the halo cuspiness, i.e., increased 
central mass concentration and progressively smaller size of the halo core. 
Figs.~3 and 6 of ES02 demonstrate that the regular (stable) region shrinks from 
$\sim 5-6$~kpc in RT\,1 to about 0.5~kpc in RT\,3, along with the increased
halo 
cuspiness, not leaving much space for the bar to survive. When the bar feedback 
onto the halo is fully accounted for, i.e., in live triaxial halos ---
they also lose part of their asymmetry (prolateness) $\epsilon_{\rm H}$, but 
most of it remains. This explains why bars in LT\,1 and LT\,2 models behave 
qualitatively similarly to those in rigid RT models --- all of them exhibit
the bar dissolution. 

The main difference between these models and the cuspy LT\,3 is that
cuspy halos are more structurally unstable, as we discussed above, and their 
prolateness is washed out early enough to allow for the bar to develop
in a nearly axisymmetric environment. We confirm that either the onset 
of chaos dissolves the bar (e.g., LT\,1 and LT\,2) or it destroys the halo 
prolateness (e.g., LT\,3). We strongly suspect that this latter process of
is also at work in Curir et al. (2005) simulations of cosmological halos,
allowing for long-lived bars. Therefore, to some
extent long-lived bars are incompatible with the high (equatorial)
asymmetry in the DM halos. Even the mild asymmetry observed in some halos of 
nearby galaxies is sufficient to shorten the bar lifetime to $\ltorder 5$~Gyrs.

This provides an interesting constraint on halo shapes, when taken in tandem
with recent observational results. Using $HST$-based morphologies and accurate
redshifts from the Galaxy Evolution from Morphology and SEDs (GEMS; Rix et al.
2004) survey, Jogee et al. (2004, hereafter J04; 2005) recently showed that the
optical fraction and distribution of structural properties for bars with
moderate-to-high strength ($\epsilon\gtorder 0.35$) remain similar from the 
present-day, out to look-back times of $2-8$~Gyr ($z\sim 0.2-1.0$). J04 argue 
that these findings imply that 
on average bars have a long lifetime, well in excess of 2~Gyr.
A constant optical bar fraction out to $z\sim 1$ is also reported 
independently from a smaller survey by Elmegreen et al. (2004). 
The simulations in this paper, when combined with these empirical results 
on a relatively constant fraction of bars out to $z\sim 1$  and an inferred
long bar lifetime, put a lower limit on the halo equatorial
axial ratio $\beta=b/a$  of 0.9  (in potential axes) at the redshift range 
of $z\sim 0-1$. This corresponds approximately to ($b/a$)$_\rho\sim 0.75-0.8$ 
axial ratio in {\it density} distribution.
 
The disappearance of triaxial halos with $\epsilon_{\rm H}\gtorder 0.1$ in disk
galaxies at $z\ltorder 1$ seems as a corollary to our numerical simulations, 
when supplemented with the above observational results.
As discussed  in section~1, while cosmological simulations of dissipationless  
CDM galactic halos invariably produce triaxial halos, 
the triaxiality of the halo can be subsequently diluted by the addition  
of baryonic components. This is likely to happen, for instance, during the 
early formation and development 
of a galactic disk. Furthermore, this trend will be supported by a general
decrease in the frequency of galaxy interactions and mergers below $z\sim 2$.

We note the following caveat: the halo prolateness in our modeled halos is
(partially or fully) washed out by the action of the bar which leads
to a dramatic increase in the fraction of chaotic orbits --- these
cannot support triaxial figures. 
It is the inner bar-forming part of the system where chaotic orbits
dominate in asymmetric halos. In principle, we can envision the situation
when the halo figure tumbles and trajectories which originate in its
outer part cannot penetrate deep enough toward the central regions where
they are typically destabilized. Such a halo will be more stable in preserving
its prolateness, at least in its outer part. It is unclear how fast the halo 
must be tumbling for this effect to take place. Such models are beyond the 
scope of this work.  

Finally, we find it important to address the question to what extent the 
evolution of stellar bars in our $N$-body potentials of mildly triaxial halos
is {\it not} a numerical artifact, i.e., not a process driven by a numerical
diffusion. In other words, can a Poisson noise, associated with the
discreteness of the $N$-body potential, be primarily responsible for the
observed model bar dissolution, as opposed to the chaotic behavior triggered
by the competing forces from the triaxial halo and a bar in ES02 for smooth 
analytical potentials? One can argue that that the graininess
of the $N$-body potential is identical in the axisymmetric and triaxial
halo models. Because there is no indication that this potential has caused any
numerical `relaxation' effects in the former models, the latter ones
will not be dominated by these effects as well. In principle, the 
consequences of the discreteness noise in the `mixed' (triaxial halo/bar)
models, i.e., models which are intrinsically more chaotic, can be more complex 
than in the axisymmetric models. Nevertheless, albeit indirectly, this
supports our point of view that it is the chaotic dynamics which drives the
evolution of stellar bar in triaxial halos modeled here.

To provide the most direct answer to this question is to
analyze the numerical models presented here in terms of Lyapunov exponents
of ES02, but this is outside the scope of our work and will be addressed
elsewhere. However, even if this approach will be taken, it is not without 
caveats as we deal here with a transient  and not a classical chaos which
is defined in the asymptotic limit and corresponds to an exponential 
divergence of a trajectory in a {\it fixed} potential (e.g., Lichtenberg \&
Lieberman 1995).

In summary, we find that even a mild halo triaxiality of 
$\sim 0.9$ in potential axial ratio, dissolves the stellar bar on a 
timescale of $\ltorder 5$~Gyr, sometimes leaving behind a weak oval distortion. 
Especially in the live triaxial halos, as long as a bar is present,
strong spiral structure develop outside the bar and its strength
depends on the mutual orientation of bars and the halo major axis.
In comparable live axisymmetric halos, there is limited secular evolution,  
either growth or decay, of the embedded bars.  In  these cases, 
the halo core sizes correlate directly with the bar sizes and their 
central mass concentration --- with the bar pattern speeds.
Cuspy halos are more susceptible to washing out of their triaxiality due 
to the action of the bar, and the subsequent evolution is similar to that  
of axisymmetric halos, where the bar does survive.
We have interpreted the bar evolution in live asymmetric  halos, as well
as in mildly triaxial and cuspy models in terms of the orbital structure, 
the  development of chaos and the feedback between the halo, disk and bar.
We find that damping of the bar instability in such halos puts a tight 
upper limit on halo prolateness in {\it disk} galaxies in the range of 
redshifts extending from the
local Universe to $z\sim 1$, in the light of recent results on a constant
optical fraction of bar from the present-day out to these epochs.

\acknowledgments
We are grateful to Lia Athanassoula, Amr El-Zant, Clayton Heller and Inma 
Martinez-Valpuesta for numerous discussions. This research has been partially 
supported by NASA/LTSA grant NAG 5-13063 (S.J. and I.S.), NASA/ATP NAG 5-10823, 
HST AR-09546.01-A and 10284, and NSF grant AST-0206251 (I.S.). Simulations 
have been performed on a dedicated LinuX Cluster and we thank Brian Doyle for 
technical support.

\end{document}